\documentclass[usenatbib]{mn2e}

\usepackage{rotating,pdflscape}
\usepackage{graphicx}
\usepackage{amssymb,amsmath}
\usepackage{natbib}
\usepackage[top= 2cm,bottom=2cm,left=2cm,right=2cm]{geometry}

\begin{document}
\title{The Spitzer Local Volume Legacy (LVL) Global Optical Photometry}

\author[D. Cook et al.]
{David O. Cook,$^1$
Daniel A. Dale$^1$
Benjamin D. Johnson$^2$
Liese Van Zee$^3$
\newauthor
Janice C. Lee$^4$
Robert C. Kennicutt$^{5,6}$
Daniela Calzetti$^7$
Shawn M. Staudaher$^1$
\newauthor
Charles W. Engelbracht$^{8}$\\
$^1$Department of Physics \& Astronomy, University of Wyoming, Laramie, WY 82071, USA; dcook12$@$uwyo.edu\\
$^2$UPMC-CNRS, UMR7095, Institut d'Astrophysique de Paris, F-75014, Paris, France\\
$^3$Department of Astronomy, Indiana University, Bloomington, IN 47405, USA\\
$^4$Space Telescope Science Institute, 3700 San Martin Drive, Baltimore, MD 21218, USA\\
$^5$Institute of Astronomy, University of Cambridge, Cambridge CB3 0HA, UK\\
$^6$Steward Observatory, University of Arizona, Tucson, AZ 85721, USA\\
$^7$Department of Astronomy, University of Massachusetts, Amherst, MA 01003, USA\\
$^8$Raytheon Company, 1151 East Hermans Road, Tucson, AZ 85756, USA; Deceased}

\maketitle
\begin{abstract}
We present the global optical photometry of 246 galaxies in the Local Volume Legacy (LVL) survey. The full volume-limited sample consists of 258 nearby ($D<$11~Mpc) galaxies whose absolute $B-$band magnitude span a range of $-9.6 < M_B < -20.7$ mag. A composite optical ($UBVR$) data set is constructed from observed $UBVR$ and SDSS $ugriz$ imaging, where the $ugriz$ magnitudes are transformed into $UBVR$. We present photometry within three galaxy apertures defined at UV, optical, and IR wavelengths. Flux comparisons between these apertures reveal that the traditional optical R25 galaxy apertures do not fully encompass extended sources. Using the larger IR apertures we find color-color relationships where later-type spiral and irregular galaxies tend to be bluer than earlier-type galaxies. These data provide the missing optical emission from which future LVL studies can construct the full panchromatic (UV-optical-IR) spectral energy distributions.
\end{abstract}


\section{Introduction}
The Local Volume Legacy (LVL) survey has provided large data sets which include ultraviolet \citep[UV;][]{lee11} non-ionizing stellar continuum, H$\alpha$ nebular emission \citep{kennicutt08}, and infrared \citep[IR;][]{dale09} dust emission of hundreds of local galaxies. Due to the volume-limited nature of LVL, the galaxy sample consists of large spirals and a few elliptical galaxies, but mostly low-mass dwarf and irregular galaxies where 52\% of the sample has a stellar mass less than $10^9 $M$_{\odot}$ \citep{dale09}. These data have produced insights into the extreme environments of low-mass galaxies.

LVL IR studies have established and characterized monochromatic IR star formation rate indicators which break down for low luminosity galaxies due to decreased amount of dust \citet{calzetti10}. Furthermore, \citet{dale09} observed large scatter in the mid-to-total IR flux ratios for low-mass galaxies and that they exhibit low IR/UV flux ratios compared to star burst galaxies. These results indicate that the reprocessing of UV photons by dust in low-mass galaxies is different than in normal or starbursting galaxies. 

The study of \citep{lee09b} utilized the LVL H$\alpha$ and UV catalogs to compare these two star formation rate tracers and found a discrepancy in lower mass systems \citep[see also,][]{meurer09}. The underprediction of the H$\alpha$ star formation rate in low-mass galaxies has consequences on a wide range of fundamental galaxy properties (e.g., IMF, stellar models, etc.) and has given rise to a new paradigm of research investigations \citep[i.e. stochastic star formation; ][]{barnes11,slug,fumagalli11,cook12,dasilva12,eldridge12,koda12,weisz12,andrews13,hermanowicz13,barnes13,bauer13,dasilva14}.

These previous LVL studies have provided information on the star formation and interstellar dust properties of local galaxies, but the full spectral energy distribution (SED) is incomplete. The optical light significantly contributes to the overall emission of a galaxy, and therefore provides invaluable constraints for modeling a galaxy's full SED. In addition, due to the fewer technical obstacles involved in ground-based optical observations, examining relationships between optical properties and properties derived from either UV or IR is a practical and useful endeavor.

This paper presents the global photometry of the LVL optical observations. The full panchromatic (UV-optical-IR) SEDs are constructed and comparisons between optical and physical properties are explored in a follow-up study (Cook et al. 2014c; submitted).

\section{Sample \label{sec:sample}}
The LVL sample consists of 258 of our nearest galaxy neighbors reflecting a statistically complete, representative sample of the local universe. The sample selection and description are detailed in \citet{dale09}, but we provide a brief overview here. 

\begin{table*}
{General Galaxy Properties}\\
\begin{tabular}{lcccccc}
\hline
\hline
Galaxy & RA        & DEC        & $D$   & $T$ & $E(B-V)$ & $M_{B}$       \\
Name   & (J2000.0) & (J2000.0)  & (Mpc) &     & (mag)    & (AB mag)      \\
(1)    & (2)       & (3)        & (4)   & (5) & (6)      & (7)           \\
\hline
                 WLM   & 00:01:58.16   & $-$15:27:39.3   &   ~0.92   &     ~10   &   0.04   &     -13.56   \\ 
             NGC0024   & 00:09:56.54   & $-$24:57:47.3   &   ~8.13   &     ~~5   &   0.02   &     -17.46   \\ 
             NGC0045   & 00:14:03.99   & $-$23:10:55.5   &   ~7.07   &     ~~8   &   0.02   &     -18.19   \\ 
         NGC0055\dag   & 00:14:53.60   & $-$39:11:47.9   &   ~2.17   &     ~~9   &   0.01   &     \ldots   \\ 
             NGC0059   & 00:15:25.13   & $-$21:26:39.8   &   ~5.30   &    $-$3   &   0.02   &     -15.62   \\ 
         ESO410-G005   & 00:15:31.56   & $-$32:10:47.8   &   ~1.90   &    $-$1   &   0.01   &     -11.34   \\ 
        SCULPTOR-DE1   & 00:23:51.70   & $-$24:42:18.0   &   ~4.20   &     ~10   &   0.02   &     -11.00   \\ 
         ESO294-G010   & 00:26:33.37   & $-$41:51:19.1   &   ~1.90   &    $-$3   &   0.01   &     -11.09   \\ 
              IC1574   & 00:43:03.82   & $-$22:14:48.8   &   ~4.92   &     ~10   &   0.02   &     -14.05   \\ 
             NGC0247   & 00:47:08.55   & $-$20:45:37.4   &   ~3.65   &     ~~7   &   0.02   &     -18.31   \\ 
\hline
\end{tabular} \\
\caption{Column 1: Galaxy name; galaxies where the photometry has been removed is denoted with a \dag~symbol. Column 2 and 3: J2000 right ascension and declination used for centering the R25 apertures of Table 2 which are based on the RC3 catalog values. Column 4: distance in Mpc from \citet{kennicutt08}. Column 5: RC3 Morphological T-type from \citet{kennicutt08}. Column 6: Milky Way foreground extinction from \citet{schlegel98}. Column 7: absolute $B-$band magnitude derived from the apparent magnitude of this study using the distance of column 4; these magnitudes have been corrected for Milky Way foreground extinction using column 6 and the reddening curve of \citet{draine03}. The full table is available online.}

\label{tab:genprop}
\end{table*}

The LVL sample was built upon the samples of two previous nearby galaxy surveys: ACS Nearby Galaxy Survey Treasury \citep[ANGST;][]{dalcanton09} and 11~Mpc H$\alpha$ and Ultraviolet Galaxy Survy \citep[11 HUGS;][]{kennicutt08,lee11}. The final LVL sample consists of galaxies that appear outside the Galactic plane ($|b| > 20^{\circ}$), have a distance less than 11~Mpc ($D \leq$ 11~Mpc), span an absolute $B-$band magnitude of $-9.6 < M_B < -20.7$, and span an RC3 catalog galaxy type range of $-5 < T < 10$. Although the galaxy morphology composition is diverse, the LVL sample is dominated by dwarf galaxies due to its volume-limited nature. The full LVL sample and basic properties are listed in Table~\ref{tab:genprop}.


\section{Data \label{sec:data}}
In this section we describe the optical observations presented in this study. Furthermore, we describe procedures for removal of contaminating sources, define three elliptical apertures used for photometry, and present the subsequent optical photometry within these apertures. 

\subsection{UBVR Data \label{sec:ubvr}}
We have obtained $UBVR$ ground-based data from 1--2 meter class telescopes located at Cerro Tololo Inter-American Observatory (CTIO), Kitt Peak National Observatory (KPNO), Vatican Advanced Technology Telescope (VATT), and the Wyoming Infra-Red Observatory (WIRO). 

Although each telescope has its own characteristics, the collection of optical images has a median pixel scale of 0\farcs5 per pix with a standard deviation of 0\farcs1 per pix and a median seeing of 1\farcs4 with a standard deviation of 0\farcs4. In addition, the median exposure times for the $UBVR$ filters are 2700, 1800, 1260, and 840 seconds, respectively.

All data reduction was performed with standard IRAF tasks. Each image was bias- and dark- subtracted, and flat-fielded. Dome flats were used only when skyflats were not available. Calibration stars were taken several times a night and were used to correct fluxes to zero airmass, calculate photometric zero points, and calculate first-order color corrections.

The 5$\sigma$ limiting magnitude for a point source using the measured seeing as the aperture (2$\times$FWHM of the PSF) was calculated via the standard deviation of sky values within several sky regions selected to avoid bright stars. The median 5$\sigma$ limiting magnitudes for all images are 22.8, 23.6, 23.1, and 22.6 mag with a standard deviation of 0.77, 0.79, 0.71, 0.78 mag for $UBVR$, respectively.

\subsection{SDSS Data \label{sec:ugriz}}
We have also downloaded SDSS $ugriz$ imaging taken from SDSS DR7 \citep[Sloan Digital Sky Survey;][]{sdss7}. For each galaxy, mosaic images were constructed using the utility SWARP \citep{swarp}, sky-subtracted, and photometrically calibrated to the AB magnitude system; note that the SDSS magnitudes are corrected for AB magnitude offsets as prescribed by the DR7 data release website.\footnote{http://www.sdss.org/dr7/algorithms/fluxcal.html} 

The 5$\sigma$ limiting magnitudes of the SDSS mosaics were calculated in an identical manner to that of our $UBVR$ data resulting in point source detection limits of 21.6, 22.8, 22.4, 21.9, 20.4 mag with a standard deviation of 0.23, 0.20, 0.17, 0.17, 0.21 mag for $ugriz$, respectively. These limiting magnitudes are in good agreement with the 95\% completeness limits quoted in the SDSS DR7 documentation. 




\begin{figure*}
  \begin{center}
  \includegraphics[scale=0.8]{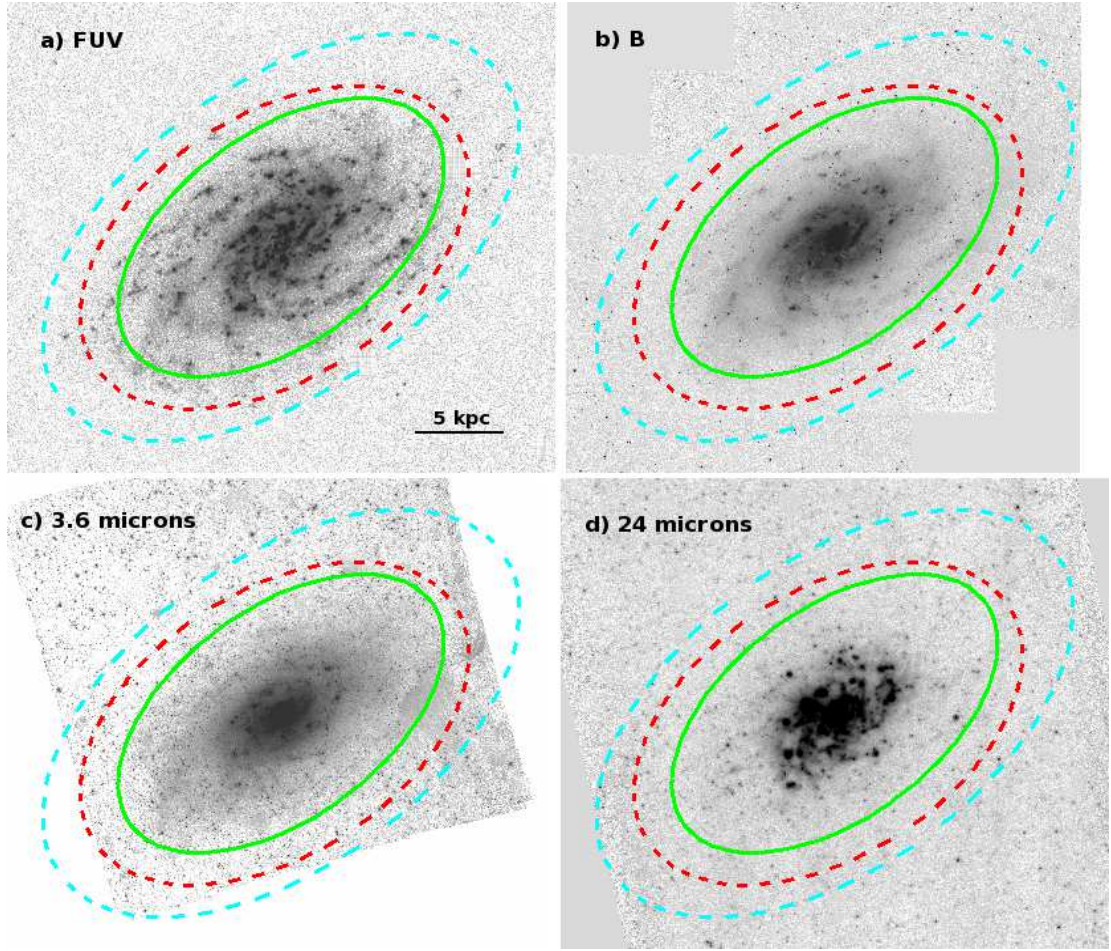}
  \caption{Inverted images for the LVL spiral galaxy NGC2403 in 4 filters: FUV, B, 3.6$\mu m$, and 24$\mu m$, for panels a, b, c, and d, respectively. We perform photometry in the 3 apertures overlaid in each panel: optical R25 (green-solid line), IR \citep[red-dashed line;][]{dale09}, and UV \citep[cyan-dashed line;][]{lee11}. To make direct comparisons across UV--IR we present optical fluxes within the IR apertures. The IR apertures have been visually checked to ensure that they are adequately large to encompass all of the optical and typically all of the UV and IR emission.}
   \label{fig:2403}
   \end{center}
\end{figure*}  

\begin{figure*}
  \begin{center}
  \includegraphics[scale=0.8]{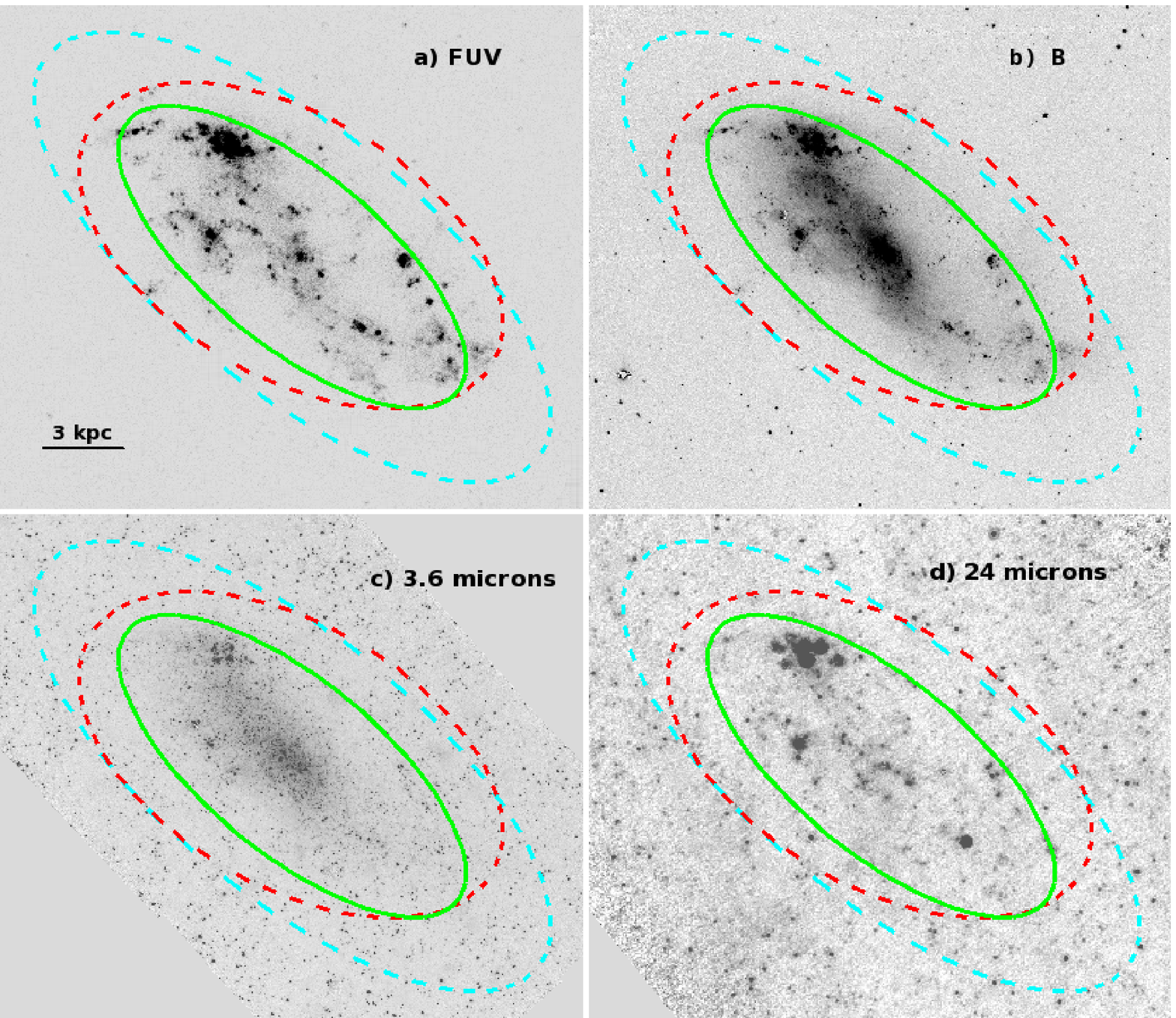}
  \caption{Inverted images of the LVL dwarf galaxy UGC05666 where the panels and ellipses are the same as Figure~\ref{fig:2403}.}
   \label{fig:5666}
   \end{center}
\end{figure*}  

\subsection{Removal of Contaminating Sources \label{sec:remove}}
To prepare the optical data for photometry, contaminating sources (e.g., background galaxies and foreground stars) were identified and masked within the galaxy apertures studied here (see \S\ref{sec:aps}). We utilize the contaminant regions identified by \citet{dale09} as an initial catalog. These regions were identified based on Spitzer Space Telescope IR colors and high resolution $HST$ images when archival $HST$ images were available (i.e., not available for all LVL galaxies.). These contaminant regions were overlaid onto each optical image, visually inspected, and the contaminant region sizes were adjusted to account for the relative apparent brightness of each source. 

Although the work of \citet{dale09} compared the LVL IR images to high resolution $HST$ images, contaminating optical sources are not always detectable at IR wavelengths and hence may not have been recorded as contaminating sources. Furthermore, new $HST$ observations since the publication date of \citet{dale09} are available to identify additional contaminating sources. For these reasons, we have compared our optical data to all available $HST$ single-band images and multi-band color images via the Hubble Legacy Archive\footnote{http://hla.stsci.edu/hlaview.html}. $HST$ images were visually checked for sources with spiral structure, extended profiles, and  optical colors indicating a background galaxy, and sources with diffraction spikes indicating a foreground star. The new contaminant regions were tailored to each optical image and added to the modified contamination source catalog of \citet{dale09}. With an updated contaminating source catalog, each contaminant was removed through an interpolation of the surrounding local sky using the IRAF task IMEDIT. The region files which define these contaminating sources are available online or by request.

\subsection{Galaxy Apertures \label{sec:aps}}
We perform global photometry on the LVL optical images within three apertures defined at UV, optical, and IR wavelengths. The optical apertures are taken from the RC3 catalog \citep{rc3} as tabulated by the VizieR catalog,\footnote{http://vizier.u-strasbg.fr/} and are defined as isophotal ellipses with a surface brightness of 25 mag/arcsec$^2$ in the $B-$band filter. The optical apertures are listed in Table~\ref{tab:apr25}. The other two apertures were chosen to provide direct flux comparisons with two previous LVL studies which have defined separate elliptical apertures.

\begin{sidewaystable}
\vspace{7cm}
\begin{center}
{Photometry Within R25 Apertures}
\end{center}
\footnotesize
\begin{tabular}{lcccccccccccc}
\hline
\hline
Galaxy & a                    & b                    & P.A.  & $U$      & $B$      & $V$      & $R$      & $u$      & $g$      & $r$      & $i$      & $z$ \\ 
       & ($^{\prime\prime}$)  & ($^{\prime\prime}$)  & (deg) & (AB mag) & (AB mag) & (AB mag) & (AB mag) & (AB mag) & (AB mag) & (AB mag) & (AB mag) & (AB mag) \\
(1)    & (2)                  & (3)                  & (4)   & (5)      & (6)      & (7)      & (8)      & (9)      & (10)     & (11)     & (12)     & (13) \\
\hline
                 WLM   &  ~~~344.4   &  ~~~119.4   &     ~~4.0   &                    \ldots   &            11.50$\pm$0.03   &            11.10$\pm$0.03   &            10.93$\pm$0.03   &                    \ldots   &                    \ldots   &                    \ldots   &                    \ldots   &                    \ldots  \\ 
             NGC0024   &  ~~~172.6   &  ~~~~40.5   &     ~46.0   &                    \ldots   &            12.20$\pm$0.03   &            11.58$\pm$0.03   &            11.46$\pm$0.03   &                    \ldots   &                    \ldots   &                    \ldots   &                    \ldots   &                    \ldots  \\ 
             NGC0045   &  ~~~255.3   &  ~~~176.7   &     142.0   &            11.92$\pm$0.03   &            11.21$\pm$0.03   &            10.73$\pm$0.03   &            10.58$\pm$0.03   &                    \ldots   &                    \ldots   &                    \ldots   &                    \ldots   &                    \ldots  \\ 
         NGC0055\dag   &  ~~~970.8   &  ~~~168.7   &     108.0   &                    \ldots   &                    \ldots   &                    \ldots   &                    \ldots   &                    \ldots   &                    \ldots   &                    \ldots   &                    \ldots   &                    \ldots  \\ 
             NGC0059   &  ~~~~78.9   &  ~~~~39.5   &     127.0   &            14.01$\pm$0.03   &            13.21$\pm$0.03   &            12.59$\pm$0.03   &            12.31$\pm$0.03   &                    \ldots   &                    \ldots   &                    \ldots   &                    \ldots   &                    \ldots  \\ 
         ESO410-G005   &  ~~~~39.5   &  ~~~~31.4   &     ~54.0   &                    \ldots   &            15.32$\pm$0.03   &                    \ldots   &            14.58$\pm$0.03   &                    \ldots   &                    \ldots   &                    \ldots   &                    \ldots   &                    \ldots  \\ 
        SCULPTOR-DE1   & ~~~\ldots   & ~~~\ldots   &   ~\ldots   &                    \ldots   &                    \ldots   &                    \ldots   &                    \ldots   &                    \ldots   &                    \ldots   &                    \ldots   &                    \ldots   &                    \ldots  \\ 
         ESO294-G010   &  ~~~~32.9   &  ~~~~21.2   &     ~~6.0   &                    \ldots   &            15.76$\pm$0.03   &                    \ldots   &            14.97$\pm$0.03   &                    \ldots   &                    \ldots   &                    \ldots   &                    \ldots   &                    \ldots  \\ 
              IC1574   &  ~~~~64.1   &  ~~~~23.3   &     175.0   &            15.25$\pm$0.03   &            14.73$\pm$0.03   &            14.35$\pm$0.03   &            14.24$\pm$0.03   &            15.29$\pm$0.03   &            14.52$\pm$0.03   &            14.20$\pm$0.03   &            14.04$\pm$0.03   &            14.11$\pm$0.03  \\ 
             NGC0247   &  ~~~641.4   &  ~~~207.5   &     174.0   &            10.11$\pm$0.03   &            ~9.61$\pm$0.03   &            ~9.10$\pm$0.03   &            ~8.87$\pm$0.03   &                    \ldots   &                    \ldots   &                    \ldots   &                    \ldots   &                    \ldots  \\ 
\hline
\end{tabular} \\
\begin{rotate}{270}
\rotcaption{The LVL global photometry within the optical-R25 apertures. Column 1: The galaxy name; galaxies where the photometry has been removed is denoted with a \dag~symbol. Columns 2 and 3: The semi-major and semi-minor axis of the the R25 apertures in units of arcseconds. Column 4: The position angle of the R25 apertures in units of degrees measured east of north. Columns 5, 6, 7, and 8: The $UBVR$ magnitudes of each galaxy with errors. Columns 9, 10, 11, 12, and 13: The $ugriz$ magnitudes of each galaxy with errors. The full table is available online.}

\label{tab:apr25}
\end{rotate}
\end{sidewaystable}

\begin{sidewaystable}
\vspace{6cm}
\begin{center}
{Photometry Within the IR Apertures of \citet{dale09}}
\end{center}
\footnotesize
\begin{tabular}{lcccccccccccc}
\hline
\hline
Galaxy & a                    & b                    & P.A.  & $U$      & $B$      & $V$      & $R$      & $u$      & $g$      & $r$      & $i$      & $z$ \\ 
       & ($^{\prime\prime}$)  & ($^{\prime\prime}$)  & (deg) & (AB mag) & (AB mag) & (AB mag) & (AB mag) & (AB mag) & (AB mag) & (AB mag) & (AB mag) & (AB mag) \\
(1)    & (2)                  & (3)                  & (4)   & (5)      & (6)      & (7)      & (8)      & (9)      & (10)     & (11)     & (12)     & (13) \\
\hline
                 WLM   &   ~~336.0   &   ~~169.5   &    ~~~0.0   &                    \ldots   &            11.40$\pm$0.03   &            10.99$\pm$0.03   &            10.80$\pm$0.03   &                    \ldots   &                    \ldots   &                    \ldots   &                    \ldots   &                    \ldots  \\ 
             NGC0024   &   ~~150.5   &   ~~108.0   &     225.0   &                    \ldots   &            12.16$\pm$0.03   &            11.54$\pm$0.03   &            11.43$\pm$0.03   &                    \ldots   &                    \ldots   &                    \ldots   &                    \ldots   &                    \ldots  \\ 
             NGC0045   &   ~~288.5   &   ~~228.0   &     336.0   &            11.88$\pm$0.03   &            11.13$\pm$0.03   &            10.65$\pm$0.03   &            10.51$\pm$0.03   &                    \ldots   &                    \ldots   &                    \ldots   &                    \ldots   &                    \ldots  \\ 
         NGC0055\dag   &    1125.5   &   ~~357.0   &     106.0   &                    \ldots   &                    \ldots   &                    \ldots   &                    \ldots   &                    \ldots   &                    \ldots   &                    \ldots   &                    \ldots   &                    \ldots  \\ 
             NGC0059   &   ~~127.5   &   ~~~89.5   &     301.0   &            13.98$\pm$0.03   &            13.08$\pm$0.03   &            12.45$\pm$0.03   &            12.17$\pm$0.03   &                    \ldots   &                    \ldots   &                    \ldots   &                    \ldots   &                    \ldots  \\ 
         ESO410-G005   &   ~~~60.5   &   ~~~45.0   &     307.0   &                    \ldots   &            15.11$\pm$0.03   &                    \ldots   &            14.34$\pm$0.03   &                    \ldots   &                    \ldots   &                    \ldots   &                    \ldots   &                    \ldots  \\ 
        SCULPTOR-DE1   &   ~~~80.0   &   ~~~51.5   &    ~~~0.0   &                    \ldots   &            17.01$\pm$0.04   &                    \ldots   &            16.26$\pm$0.03   &                    \ldots   &                    \ldots   &                    \ldots   &                    \ldots   &                    \ldots  \\ 
         ESO294-G010   &   ~~~82.0   &   ~~~50.5   &    ~~~0.0   &                    \ldots   &            15.33$\pm$0.03   &                    \ldots   &            14.51$\pm$0.03   &                    \ldots   &                    \ldots   &                    \ldots   &                    \ldots   &                    \ldots  \\ 
              IC1574   &   ~~101.0   &   ~~~61.5   &    ~~~0.0   &            14.96$\pm$0.03   &            14.47$\pm$0.03   &            14.03$\pm$0.03   &            13.95$\pm$0.03   &            14.77$\pm$0.04   &            14.28$\pm$0.03   &            13.89$\pm$0.03   &            13.66$\pm$0.03   &            13.84$\pm$0.04  \\ 
             NGC0247   &   ~~738.0   &   ~~290.5   &     352.0   &            10.07$\pm$0.03   &            ~9.56$\pm$0.03   &            ~9.05$\pm$0.03   &            ~8.81$\pm$0.03   &                    \ldots   &                    \ldots   &                    \ldots   &                    \ldots   &                    \ldots  \\ 
\hline
\end{tabular} \\
\begin{rotate}{270}
\rotcaption{The LVL global photometry within the IR apertures defined by \citet{dale09}. Column 1: The galaxy name; galaxies where the photometry has been removed is denoted with a \dag~symbol. Columns 2 and 3: The semi-major and semi-minor axis of the the IR apertures in units of arcseconds. Column 4: The position angle of the IR apertures in units of degrees measured east of north. Columns 5, 6, 7, and 8: The $UBVR$ magnitudes of each galaxy with errors. Columns 9, 10, 11, 12, and 13: The $ugriz$ magnitudes of each galaxy with errors. The full table is available online.}

\label{tab:apir}
\end{rotate}
\end{sidewaystable}

\begin{sidewaystable}
\vspace{-10cm}
\begin{center}
{Photometry Within the UV Apertures of \citet{lee11}}
\end{center}
\footnotesize
\begin{tabular}{lcccccccccccc}
\hline
\hline
Galaxy & a                    & b                    & P.A.  & $U$      & $B$      & $V$      & $R$      & $u$      & $g$      & $r$      & $i$      & $z$ \\ 
       & ($^{\prime\prime}$)  & ($^{\prime\prime}$)  & (deg) & (AB mag) & (AB mag) & (AB mag) & (AB mag) & (AB mag) & (AB mag) & (AB mag) & (AB mag) & (AB mag) \\
(1)    & (2)                  & (3)                  & (4)   & (5)      & (6)      & (7)      & (8)      & (9)      & (10)     & (11)     & (12)     & (13) \\
\hline
                 WLM   &   ~~516.0   &   ~~180.0   &     ~~4.0   &                    \ldots   &            11.30$\pm$0.03   &            10.87$\pm$0.03   &            10.66$\pm$0.03   &                    \ldots   &                    \ldots   &                    \ldots   &                    \ldots   &                    \ldots  \\ 
             NGC0024   &   ~~258.0   &   ~~~58.0   &     ~46.0   &                    \ldots   &            12.12$\pm$0.03   &            11.50$\pm$0.03   &            11.39$\pm$0.03   &                    \ldots   &                    \ldots   &                    \ldots   &                    \ldots   &                    \ldots  \\ 
             NGC0045   &   ~~378.0   &   ~~262.0   &     -38.0   &            11.84$\pm$0.03   &            11.07$\pm$0.03   &            10.60$\pm$0.03   &            10.44$\pm$0.03   &                    \ldots   &                    \ldots   &                    \ldots   &                    \ldots   &                    \ldots  \\ 
         NGC0055\dag   &    1452.0   &   ~~251.0   &     -72.0   &                    \ldots   &                    \ldots   &                    \ldots   &                    \ldots   &                    \ldots   &                    \ldots   &                    \ldots   &                    \ldots   &                    \ldots  \\ 
             NGC0059   &   ~~150.0   &   ~~~75.0   &     -53.0   &            13.98$\pm$0.03   &            13.08$\pm$0.03   &            12.45$\pm$0.03   &            12.16$\pm$0.03   &                    \ldots   &                    \ldots   &                    \ldots   &                    \ldots   &                    \ldots  \\ 
         ESO410-G005   &   ~~108.0   &   ~~~83.0   &     ~54.0   &                    \ldots   &            14.88$\pm$0.03   &                    \ldots   &            14.08$\pm$0.03   &                    \ldots   &                    \ldots   &                    \ldots   &                    \ldots   &                    \ldots  \\ 
        SCULPTOR-DE1   &   ~~102.0   &   ~~~79.0   &     -10.0   &                    \ldots   &            16.75$\pm$0.04   &                    \ldots   &            15.97$\pm$0.03   &                    \ldots   &                    \ldots   &                    \ldots   &                    \ldots   &                    \ldots  \\ 
         ESO294-G010   &   ~~~72.0   &   ~~~46.0   &     ~~6.0   &                    \ldots   &            15.35$\pm$0.03   &                    \ldots   &            14.53$\pm$0.03   &                    \ldots   &                    \ldots   &                    \ldots   &                    \ldots   &                    \ldots  \\ 
              IC1574   &   ~~120.0   &   ~~~46.0   &      -5.0   &            14.97$\pm$0.03   &            14.47$\pm$0.03   &            14.04$\pm$0.03   &            13.94$\pm$0.03   &            14.80$\pm$0.04   &            14.26$\pm$0.03   &            13.89$\pm$0.03   &            13.69$\pm$0.03   &            13.82$\pm$0.04  \\ 
             NGC0247   &   ~~954.0   &   ~~307.0   &      -6.0   &            10.05$\pm$0.03   &            ~9.55$\pm$0.03   &            ~9.03$\pm$0.03   &            ~8.80$\pm$0.03   &                    \ldots   &                    \ldots   &                    \ldots   &                    \ldots   &                    \ldots  \\ 
\hline
\end{tabular} \\
\begin{rotate}{270}
\rotcaption{The LVL global photometry within the UV apertures defined by \citet{lee11}. Column 1: The galaxy name; galaxies where the photometry has been removed is denoted with a \dag~symbol. Columns 2 and 3: The semi-major and semi-minor axis of the the UV apertures in units of arcseconds. Column 4: The position angle of the UV apertures in units of degrees measured east of north. Columns 5, 6, 7, and 8: The $UBVR$ magnitudes of each galaxy with errors. Columns 9, 10, 11, 12, and 13: The $ugriz$ magnitudes of each galaxy with errors. The full table is available online.}

\label{tab:apuv}
\end{rotate}
\end{sidewaystable}

The Spitzer Space Telescope IR elliptical apertures of \citet{dale09} were chosen to encompass the majority of the emission seen at GALEX UV (1500\AA--2300\AA) and Spitzer IR (3.6$\mu$m--160$\mu$m) wavelengths. In practice, these apertures were usually determined by the extent of the 3.6$\mu$m emission given the superior sensitivity of the 3.6$\mu$m array coupled with the relatively bright emission from older stellar populations at this wavelength. However, in several instances the emission between 1500\AA--2300\AA~or at 160$\mu$m wavelengths were spatially more extended, and thus these wavelengths were used to determine the IR apertures. The resulting median ratio of IR-to-optical semi-major axes is 1.5 (see Figures \ref{fig:2403} and \ref{fig:5666}); the IR apertures are listed in Table~\ref{tab:apir}.

The GALEX UV elliptical apertures of \citet{lee11} were defined as an isophotal ellipse outside of which the photometric error was greater than 0.8 mag or the intensity fell below the sky level. The resulting median ratio of UV-to-optical semi-major axes is 2.3 (see Figures \ref{fig:2403} and \ref{fig:5666}). The UV apertures are listed in Table~\ref{tab:apuv}. 

Figures \ref{fig:2403} and \ref{fig:5666} show all three apertures overlaid onto UV, optical, and IR images for a spiral and dwarf galaxy, respectively. The dashed-cyan, solid-green, and dashed-red ellipses represent the UV, optical, and IR apertures, respectively. The IR and UV apertures are generally larger and encompass more extended features compared to the optical apertures which is most notable in panels (a) and (b) of Figure~\ref{fig:5666}. There are at least three features in panel (b) of Figure~\ref{fig:5666} which extend past the R25 aperture (more clearly seen in the UV image of panel a) but not past the UV or IR apertures. The later part of the next section shows that the global fluxes measured within the R25 apertures tend to be lower compared to the relatively similar global fluxes measured within both the UV and IR apertures.

\subsection{Photometry \label{sec:phot}}
There are six galaxies (BK06N, DDO078, IC5152, IKN, NGC0404, and UGC07242) which are removed from this analysis due to contaminating light from bright nearby stars. These bright stars are either inside or in such proximity to all three galaxy apertures that the contaminating light could not be cleanly edited or a large fraction of the galaxy's light needed to be masked in order to remove the star. The photometry of these galaxies were deemed uncertain and were subsequently removed from this analysis. Furthermore, there are five galaxies (SMC, LMC, NGC0055, NGC0300, \& NGC0598) whose optical data were not obtained due to their large angular size. In addition, no optical observations were obtained for the galaxy MCG-05-13-004. There are no fluxes listed for these 12 galaxies in Table~\ref{tab:apr25}, \ref{tab:apir}, or \ref{tab:apuv}. The remaining 246 galaxies have been observed in at least two optical filters and constitute the LVL optical sample.

Photometry was carried out via the IRAF task IMCNTS where sky apertures far from the galaxy were visually selected in each image to avoid any sources. A 3$\sigma$ clipping routine was used to calculate the final sky values. The photometric uncertainties were calculated in a manner similar to the IRAF task PHOT:
\begin{equation}
\sigma = \sqrt{f(\lambda) + A(ap)*sky^2 + \frac{A(ap)^2*sky^2}{A(sky)}},
\label{eqn:photerr}
\end{equation}

\noindent where $f(\lambda)$ is the galaxy's flux in units of counts, $A(ap)$ is the area of the galaxy's aperture in units of pixels squared, $sky$ is the standard deviation of the sky values per pixel, and $A(sky)$ is the area of the sky apertures in units of pixels squared. The first term in Equation \ref{eqn:photerr} is the Poisson error, the second term is the scatter in the sky values within the galaxy's aperture, and the third term is the uncertainty in the mean sky brightness. 

The $UBVR$ data were originally calibrated to the Vega magnitude system, but were transformed to the AB magnitude system via the prescription of \citet{blanton07}. The original Vega system calibration applied a first-order color correction. We have characterized the calibration errors (e.g., zeropoint and color term errors) of all $UBVR$ photometry at $\sim2-3$\% and have added a conservative 3\% in quadrature to the instrumental errors of Equation~\ref{eqn:photerr}.

The SDSS data are calibrated to the AB magnitude system where the SDSS DR7 documentation has characterized the calibration accuracy to be 3\%\footnote{http://www.sdss.org/dr7/algorithms/fluxcal.html}. Consequently, we have added a 3\% calibration error to the instrumental errors of Equation~\ref{eqn:photerr}. The photometry for $UBVR$ and $ugriz$ within the optical, IR, and UV apertures are listed in Tables \ref{tab:apr25}, \ref{tab:apir}, and \ref{tab:apuv}, respectively. These data have not been corrected for Galactic foreground extinction and are presented in the AB magnitude system. 

There is a single non-detection for all optical images in the SDSS $u-$band for UGCA133. The magnitude for this data point is set as the upper limit for a point source defined in \S\ref{sec:ubvr}. The $u-$band photometry listed for this galaxy in Tables~\ref{tab:apr25}, \ref{tab:apir}, and \ref{tab:apuv} has no error listed and the limiting magnitude is preceded by a ``$>$" symbol indicating an upper limit.

\begin{figure}
  \begin{center}
  \includegraphics[scale=0.5]{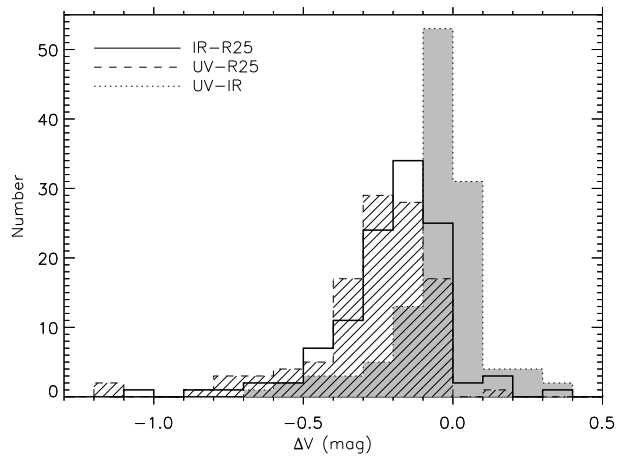}
  \caption{The $V-$band global magnitude differences measured within the IR and optical R25 apertures (IR$-$R25; unshaded), the UV and R25 apertures (UV$-$R25; shaded with diagonal lines), and the UV and IR apertures (UV$-$IR; shaded gray). The peak of both the UV$-$R25 and IR$-$R25 magnitude differences show that the UV and IR apertures encompass more flux compared to the optical R25 apertures by $0.1-0.3$ magnitudes.}
  \label{fig:compAps}
  \end{center}
\end{figure}

We also examine how the global flux of a galaxy changes when measured within different apertures (i.e., UV, optical R25, or IR). To make comparisons between the three apertures used in this study, we take the difference of the magnitudes listed in Tables~\ref{tab:apr25}, \ref{tab:apir}, and \ref{tab:apuv} and plot the resultant as histograms in Figure~\ref{fig:compAps}. However, we present only the $V-$band photometric differences since all other band-pass histograms exhibit similar overall structure. These histograms graphically present comparisons between the fluxes measured within the IR and R25 apertures (IR$-$R25; unshaded), the UV and R25 apertures (UV$-$R25; shaded with diagonal lines), and the UV and IR apertures (UV$-$IR; shaded gray). 

The peak of the IR$-$R25 and UV$-$R25 histograms are offset from a zero magnitude difference to brighter IR and UV fluxes by similar amounts ($0.1-0.3$ mag). This offset indicates that both the IR and UV apertures encompass more flux compared to the optical R25 apertures. For example, a visual inspection of panel (b) in Figure~\ref{fig:5666} shows at least three features outside of the R25 aperture (also seen in the UV image of panel a) which are contained within both the UV and IR apertures. Furthermore, the UV$-$IR histogram is peaked around a magnitude difference of zero indicating that both the UV and IR apertures encompass roughly the same amount of optical flux. This suggests that the R25 apertures, which are traditionally used to define a galaxy's spatial extent, do not fully encompass a galaxy's extended features.

\subsection{UBVR + SDSS Composite Sample \label{sec:composite}}
This study aims to compile a complete $UBVR$ data set for all galaxies; however, not all galaxies have been observed in all $UBVR$ filters. The galaxies with missing $UBVR$ observations can be supplemented with SDSS $ugriz$ observations. Table~\ref{tab:sdss} lists the number of galaxies with $UBVR$ observations in Column 2 and the number of galaxies without $UBVR$ observations but with available $ugriz$ observations in Column 3. The supplemental SDSS will require a transformation into $UBVR$ magnitudes. We transform $ugriz$ magnitudes into $UBVR$ magnitudes instead of $UBVR$ into $ugriz$ since there are more observed $UBVR$ data.

In the final composite $UBVR$ optical data set, all observed $UBVR$ data were retained and only the missing $UBVR$ data were supplemented with transformed $ugriz$ data. For example, if a galaxy has been observed in the B and R filters, the available $ugriz$ magnitudes were transformed into only the U and V magnitudes resulting in an observed BR and a transformed UV data set. Column 4 in Table~\ref{tab:sdss} presents the total number of galaxies in the composite (observed $UBVR$ plus transformed $ugriz$) LVL sample. 


Transformations were carried out via the prescription of Cook et al. (2014b; submitted). These transformation equations were derived using LVL galaxies with both $ugriz$ and $UBVR$ observations. Previous SDSS transformations derived from Landolt+Stetson calibration stars \citep[e.g.,][]{lupton04,jester05} and galaxy templates \citep{blanton07} yield increased scatter around transformation relationships and and significant offsets ($>$0.1 dex) in $UBVR$ observed-only color-color LVL trends. However, the study of Cook et al. (2014b; submitted) derives transformation relationships with significantly less scatter and no apparent offset in $UBVR$ observed-only color-color LVL trends. For a full description of the $ugriz$ to $UBVR$ transformations using galaxy-wide integrated colors see Cook et al. (2014b; submitted).

\begin{table}
\begin{center}
{LVL Composite Sample}
\begin{tabular}{ccccc}
\hline
\hline

Filter         & \# Observed   & \# SDSS       & \# Total      \\
(1)    & (2)           & (3)           & (4)           \\
\hline
   U   & 132           & ~96           & 228           \\
   B   & 185           & ~60           & 245           \\
   V   & 129           & 106           & 235           \\
   R   & 188           & ~57           & 245           \\
\hline
\end{tabular}\\
\end{center}
\caption{Explanation of galaxy sample statistics used in this study--Column2: the number of observed galaxies for each $UBVR$ filter listed in column 1; Column 3: the number of galaxies \textit{without} observed data but with SDSS data available in each filter; Column 4: the total number of galaxies in the combined observed $UBVR$ and transformed $ugriz$ data set (i.e., the sum of columns 2 and 3).}
\label{tab:sdss}
\end{table}



\section{Results} \label{sec:results} 
In this section we present properties of the $UBVR$ composite sample constructed from observed $UBVR$ and transformed SDSS $ugriz$ data. We find relationships between optical colors. 

\begin{figure}
  \begin{center}
  \includegraphics[scale=0.5]{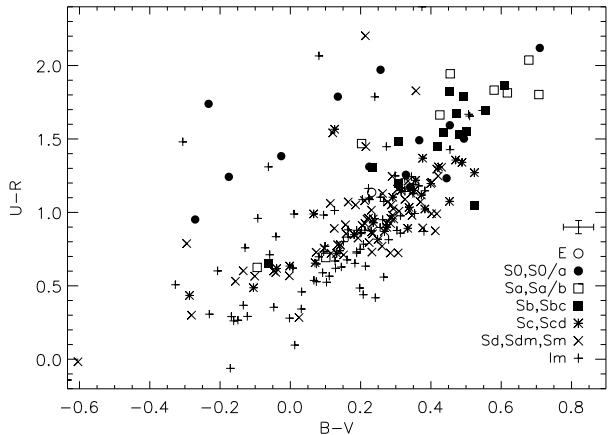}
  \caption{A color-color plot for the LVL composite $UBVR$ data, where the absolute magnitudes of each bandpass has been corrected for Milky Way foreground extinction. The error bars above the legend represent the median errors of all data plotted. There is a positive relationship with significant overlap between different galaxy types. However, earlier-type galaxies tend to be redder in color.}
  \label{fig:color1}
  \end{center}
\end{figure}  

\begin{figure}
  \begin{center}
  \includegraphics[scale=0.5]{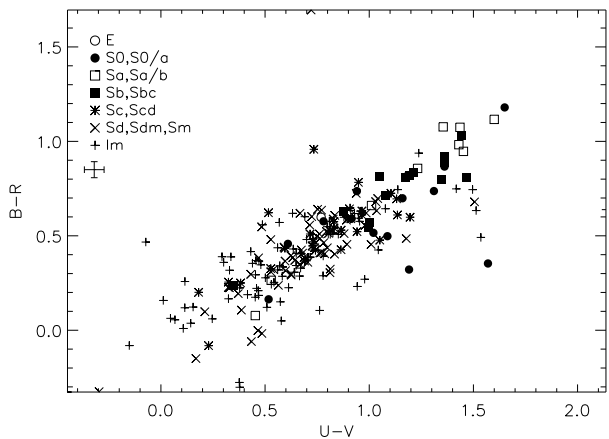}
  \caption{A color-color plot similar to that of Figure~\ref{fig:color1}, where the absolute magnitudes of each bandpass has been corrected for Milky Way foreground extinction. There is a positive relationship with significant overlap between different galaxy types. However, earlier-type galaxies tend to be redder in color.}
  \label{fig:color2}
  \end{center}
\end{figure}  

To make direct photometric comparisons we have chosen to present figures based on the fluxes within one aperture: the IR apertures of \citet{dale09}. These apertures are chosen for the following reasons: (1) the IR apertures yield similar global fluxes to those of the UV apertures, both of which exhibit greater fluxes compared to the fluxes measured within the R25 apertures (see \S\ref{sec:phot}); (2) there are fewer galaxy apertures ($N=234$) defined in the UV catalog of \citet{lee11} compared to the number of IR apertures ($N=255$) published in \citet{dale09}. The IR apertures have been visually checked to ensure that they are adequately large to encompass all of the optical emission. 

The figures in this section present absolute magnitudes. These magnitudes are corrected for Milky Way foreground extinction using the \citet{schlegel98} maps tabulated by the NASA/IPAC Extragalactic Database\footnote{http://ned.ipac.caltech.edu/} assuming the \citet{draine03} extinction curve and a reddening law of R$_V=3.1$. The $E(B-V)$ values from \citet{schlegel98} are listed in Table~\ref{tab:genprop}. 

Figures \ref{fig:color1} and \ref{fig:color2} show two optical color-color diagrams for the composite ``UBVR" LVL data set. These relationships show a general positive trend between different colors, where later-type spiral and irregular galaxies tend to be bluer than earlier-type galaxies. While this morphological trend exists in both color-color plots, there is significant overlap in the colors of different galaxy types. 
  
\section{Summary}\label{sec:summary}
This study has constructed a homogenized $UBVR$ data set for 246 LVL galaxies from newly obtained $UBVR$ imaging and compiled SDSS $ugriz$ data. We transformed $ugriz$ magnitudes into $UBVR$ magnitudes when observed $UBVR$ data were not available. Comparing the global optical fluxes between the UV apertures of \citet{lee11}, the IR apertures of \citet{dale09}, and the optical R25 apertures reveal that the traditionally defined R25 optical extent of a galaxy does not encompass the extended features of many LVL galaxies. The optical properties for the LVL sample measured within the IR apertures show relationships between different optical colors (Figures \ref{fig:color1} and \ref{fig:color2}), where earlier-type galaxies tend to show redder colors.

A follow-up publication (Cook et al. 2014c; submitted) utilizes these optical data to construct the full panchromatic (UV-optical-IR) LVL SEDs of the LVL sample. These SEDs facilitate an analysis of how physical properties (e.g., SFR, metallicity, dust, etc.) affect the observed properties of the LVL galaxy sample.

\bibliographystyle{mn2e}   
\bibliography{all}  

\end{document}